\documentclass[10pt]{article}

\usepackage{lipsum}

\usepackage{url}
\usepackage{cite}
\usepackage{amsmath,amssymb,amsfonts}
\usepackage{algorithmic}
\usepackage{graphicx}
\usepackage{textcomp}
\usepackage{xcolor}
\def\BibTeX{{\rm B\kern-.05em{\sc i\kern-.025em b}\kern-.08em
    T\kern-.1667em\lower.7ex\hbox{E}\kern-.125emX}}

\usepackage{booktabs}
\usepackage{subcaption}
\usepackage{multirow}

\usepackage[T1]{fontenc}
\usepackage{inconsolata}

\usepackage{color}
\usepackage{listings}

\usepackage[margin=1.5in]{geometry}

\definecolor{pblue}{rgb}{0.13,0.13,1}
\definecolor{pgreen}{rgb}{0,0.5,0}
\definecolor{pred}{rgb}{0.9,0,0}
\definecolor{pgrey}{rgb}{0.46,0.45,0.48}

\definecolor{mMagenta}{rgb}{0.7, 0, 0.6}

\makeatletter
\lst@Key{countblanklines}{true}[t]%
    {\lstKV@SetIf{#1}\lst@ifcountblanklines}

\lst@AddToHook{OnEmptyLine}{%
    \lst@ifnumberblanklines\else%
       \lst@ifcountblanklines\else%
         \advance\c@lstnumber-\@ne\relax%
       \fi%
    \fi}
\makeatother

\lstdefinestyle{babel} {
  emph={%
  registerMessageSerializer, registerMessageHandler,%
  registerTimerHandler, createChannel, registerChannelEventHandler, sendMessage,
  setupPeriodicTimer, cancelTimer, setupTimer, triggerNotification, openConnection,
  closeConnection, registerRequestHandler, sendRequest, sendReply, subscribeNotification,
  registerReplyHandler%
  },emphstyle={\bfseries}%
}

\begin{document}

\title{Babel: A Framework for Developing Performant and Dependable Distributed Protocols}

\author{Pedro Fouto, Pedro Ákos Costa, Nuno Preguiça, João Leitão  \\
\textit{NOVA LINCS \& DI/FCT/NOVA University of Lisbon}, \\ Lisboa, Portugal \\
\{p.fouto, pah.costa\}@campus.fct.unl.pt\\ \{nmp, jc.leitao\}@fct.unl.pt}

\maketitle

\thispagestyle{plain}
\pagestyle{plain}

\begin{abstract}

Prototyping and implementing distributed algorithms, particularly those that
address challenges related with fault-tolerance and dependability, is a time
consuming task. This is, in part, due to the need of addressing low level aspects such
as management of communication channels, controlling timeouts or periodic
tasks, and dealing with concurrency issues.
This has a significant impact for researchers that want to build prototypes for
conducting experimental evaluation; practitioners that want to compare different
design alternatives/solutions; and even for practical teaching activities on
distributed algorithms courses.

In this paper we present Babel, a novel framework to develop, implement, and
execute distributed protocols and systems.
Babel promotes an event driven programming and execution model that simplifies
the task of translating typical specifications or descriptions of algorithms
into performant prototypes, while allowing the programmer to focus on the
relevant challenges of these algorithms by transparently handling time consuming
low level aspects.
Furthermore, Babel provides, and allows the definition of, networking components
that can capture different network capabilities (e.g., P2P, Client/Server,
$\varphi$-accrual Failure Detector), making the code mostly
independent from the underlying communication aspects. Babel was built to be
generic and can be used to implement a wide variety of different classes of
distributed protocols.
We conduct our experimental work with two relevant case studies, a Peer-to-Peer
application and a State Machine Replication application, that show the generality and ease of
use of Babel and present competitive performance when compared with
significantly more complex implementations.
\end{abstract}


\section{Introduction}

Distributed systems are being increasingly used to support distinct
activities in our everyday life. With the transition to digital platforms, distributed systems
continue to grow in scale and consequently in complexity, with strict
fault-tolerance and dependability requirements. This, in turn, increases the
pressure for developing, validating, and testing different alternatives for
building dependable abstractions that support the operation of many of these
systems.


The pressure to develop better and more performant distributed systems often
requires comparing different alternatives found in the literature. This entails
creating concrete and correct implementations of solutions described in research
papers and books\,\cite{nancy,raynal2018fault,lerbook} for comparing their
performance under different conditions. This is the case for researchers
developing novel solutions that wish to compare their proposals with the
existing state of the art; for practitioners, that want to make informed
decisions regarding the design and implementation of systems under development
and operation; and even in classrooms, where students being exposed to
dependable algorithms or large-scale systems can significantly benefit from the
opportunity to implement and operate practical solutions~\cite{splay}.

%

Many (fault-tolerant) distributed algorithms/protocols are conceptually
simple to describe, with relevant examples including agreement protocols\,\cite{paxos,paxos2,paxos3},
common abstractions such as state-machine replication\,\cite{smr}, large-scale peer-to-peer
protocols\,\cite{hyparview,cyclon,peersamplingframework}, among others.
Implementing these protocols tends to be error-prone and time
consuming\,\cite{adacomponents}, negatively affecting the activities of
researchers, practitioners, and professors in the fault-tolerant and dependability
community. This is not so much due to the complexity of the algorithms
themselves, but due to the fact that such implementations require handling
several low level aspects such as the use of communication primitives,
scheduling and handling timed/periodic actions, or even deal with non-trivial concurrency
issues that naturally arise when implementing complex systems. Additionally,
stand-alone implementations require a large effort in lines of code, leading to
the logic of protocols, usually described in pseudo-code in research papers and
text books\,\cite{splay}, to be lost within the implementation details.

An alternative to practical implementation is to take advantage of simulators,
such as Peersim\,\cite{peersim} or NS2/NS3\,\cite{ns2,ns3}. However, there are
several arguments that can be made against these approaches: $i)$ such
implementations can lead to implementations that incorrectly capture a realistic
execution environment (e.g., in Peersim it is extremely easy for a node to
access the internal state of another node in the system without requiring any
communication step); $ii)$ simulators make it hard to access the performance of
competing alternatives which is many times essential for researchers and
practitioners alike; $iii)$ some simulators might require an amount of
implementation effort that is close to a real implementation (e.g., NS2 was
famous for requiring very complex protocol implementations); and $iv)$
simulators required implementing the same protocol twice (the simulator code and
the real code), without any guarantees that the real code matches the
specification of the simulator code.

Actor-based frameworks, such as Akka\,\cite{akka} or Erlang-OTP\,\cite{erlang}, offer
distributed programming abstractions that allow experienced system engineers to implement
distributed protocols quickly and efficiently. However, these high level
abstractions sometimes make it difficult to have the necessary fine-grained control
over the execution, which is fundamental for high performant implementations.
The adoption of a functional programming model also requires additional effort in translating
the specifications, usually presented in an imperative way in the literature.


In this paper we propose Babel, a novel Java framework that provides a
programming environment for implementing distributed protocols and systems.
Babel provides an event driven programming and execution model that simplifies
the task of translating typical specifications and descriptions of algorithms in
practical prototypes with no significant performance bottlenecks. As we discuss
further ahead, the abstractions provided by Babel are sufficiently generic to
enable its usage for implementing a wide variety of distributed protocols and
abstractions. Furthermore, Babel allows protocols to be implemented
independently of the networking abstractions used to support the communication
between different processes, and shields the programmer from dealing with
potentially complex concurrency aspects within, and across, distributed
protocols. This does not mean that Babel is only useful for prototyping, or that
it is non-performant. On the contrary, as we show in our evaluation, Babel was
implemented with performance in mind, allowing developers to implement
efficient, production-ready protocols. The source code for Babel is available at
\url{https://github.com/pfouto/babel-core}.

The remainder of this paper is organized as follows:
we discuss relevant related work in Section~\ref{sec:rwork}.
In Section~\ref{sec:babel}, we discuss the features and APIs of Babel.
Section~\ref{sec:casestudies} presents our experimental work with two
relevant case studies: a Peer-to-Peer application and a State Machine Replication application
using MultiPaxos\,\cite{multipaxos}.
Finally, we conclude the paper in Section~\ref{sec:end} by providing some pointers for future work.


\section{Related Work}\label{sec:rwork}
We highlight some works that share the goals of Babel of providing quick
development and reuse of practical distributed protocols and systems.



Appia\,\cite{appia} is a toolkit that takes advantage of the Java inheritance
model to allow developers to create their own protocols by extending base elements
of Appia that control the execution, which our own work also leverages.
Appia introduces the concepts of channels and sessions, which provide developers with more
control over the binding of protocols
to create different types of services within a single application however, at the cost
of requiring developers to stack protocols, which limits interactions among these different
protocols. Furthermore,
Appia has a single execution thread for executing all protocols that
compose the distributed system, which
incurs in non-negligible performance bottlenecks.

Cactus\,\cite{cactus} is a C++ framework aiming at extracting more
performance from a system and being more expressive. To this end, Cactus provides
meta-protocols that can be stacked together and that can be composed of various
smaller protocols which can be executed asynchronously. Unfortunately, Cactus does
not offer any type of concurrency management, putting that burden on the developer.


Yggdrasil\,\cite{yggdrasil} is a C framework that aims to
be lightweight and efficient. It was designed to allow the execution of distributed applications
in wireless ad-hoc scenarios, and has extensions for traditional wired IP networks.
Yggdrasil provides a clean development interface focused on the processing of four
types of events: $i)$ network messages; $ii)$ timed events; $iii)$ protocol indirect
interactions; and $iv)$ protocol direct interactions.
The last two events are used to implement message passing,
allowing for the concurrent execution of protocols while shielding the developer
from complex concurrency issues. However, Yggdrasil only supports a single
active network abstraction, meaning that it can only execute either protocols
that expect wireless ad hoc or wired IP networking in each single Yggdrasil
instance. Furthermore, being tailored for C development, it requires the
developer to be technically disciplined as to not incur in frequent invalid
memory accesses, which can lead to time consuming development.

It is also important to note that some works have proposed Java frameworks
tailored for distributed systems, but focus on the understanding of
distributed protocols. This line of work includes ViSiDia\,\cite{visidia} and
the work proposed in~\cite{javatoolkit}. Other works have focused on the
experimental evaluation and performance assessment of distributed systems in
realistic settings including SPLAY\,\cite{splay} and Kollaps\,\cite{kolaps}.
These works are complementary to our own. Whereas we focus in providing adequate
tooling for the quick, correct, and performant implementation of distributed
protocols and systems, these works simplify the task of conducting performance
assessment of such implementations.

Simulators, such as Peersim\,\cite{peersim} or NS2/NS3\,\cite{ns2,ns3}, are an
alternative for quickly developing and testing distributed systems. However,
simulators often fail to completely model a realistic execution environment and
may even allow incorrect implementations (e.g, implementations where nodes
access the internal state of other nodes). More realistic simulators tend to
require an implementation effort that is close to a real implementation (e.g.,
NS2 was famous for requiring very complex protocol implementations). In
contrast, Babel is not a simulator, but rather a framework that simplifies the
implementation of distributed systems that can be tested and deployed on real
hardware.




\section{Babel} \label{sec:babel}

\begin{figure*}[t!]
  \includegraphics[width=\textwidth]{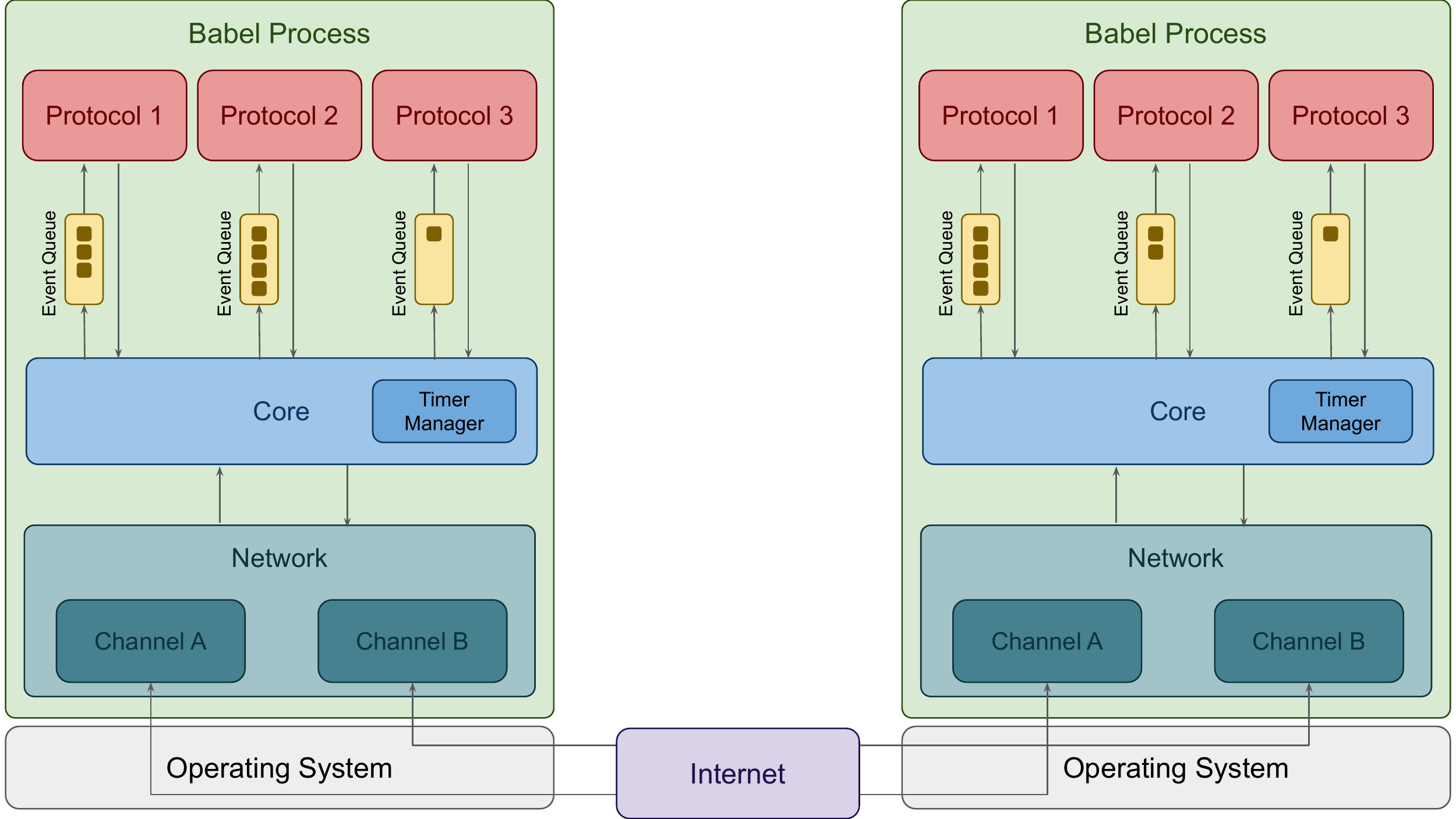}
  \caption{Architecture of Babel
  \label{fig:architecture}}
\end{figure*}

Babel is a framework that aims to simplify the development of distributed
protocols within a process that executes in real hardware.
A process can execute any number of (different) protocols that communicate
each other or/and protocols in different processes.
Babel simplifies the development by enabling the developer to focus on the logic of the protocol,
without having to deal with low level complexities associated with typical
distributed systems implementations. These complexities include interactions
among (local) protocols, handling message passing and communication aspects,
handling timers, and concurrency-control aspects within, and across, protocols
(while enabling different protocols within a process to progress independently).
Notably, Babel hides communication complexities behind abstractions called
\emph{channels} that can be extended/modified by the developer, with Babel already
offering several alternatives that capture different capabilities (e.g., P2P, Client/Server,
$\varphi$-accrual Failure Detector).
Babel is implemented in Java, taking advantage of its inheritance mechanisms,
such that developers extend abstract classes provided by the framework to
develop their own protocols and solutions. The strong typing provided by Java
allows the framework to easily enforce expected behavior, while at the same
time offering enough flexibility for the developer to implement any type of
distributed protocol or system.

Figure \ref{fig:architecture} presents the architecture of Babel. In the
example, there are two processes executing Babel, each process being composed by
three protocols and two network channels for inter-process communication.
Naturally, any distributed system operating in the real world will be composed
by more than two processes. The Babel framework in composed by three main components, which we now detail:

\subsection{Protocols} \label{sec:babel:overview:protocols} Protocols are
implemented by developers (i.e., the users of the Babel framework), and
encode all the behavior of the distributed system being designed. Each
protocol is modeled as a state machine whose state evolved by the reception
and processing of (external) events. For this purpose, each protocol contains an event
queue from which events are retrieved. These events can be \emph{Timers}, \emph{Channel Notifications}
from the network layer, \emph{Network Messages} originated from another process, or
\emph{Intra-process} events used by protocols to interact among
each other within the same process.

Each protocol is exclusively assigned a dedicated thread, which handles received
events in a serial fashion by executing their respective callbacks. In a single
Babel process, any number of protocols may be executing simultaneously, allowing
multiple protocols to cooperate (i.e., multi-threaded execution), while
shielding developers from concurrency issues, as all communication between
protocols is done via message passing.

From the developer's point-of-view, a protocol is responsible for defining the
callbacks used to process the different types of events in its queue. The
developer registers the callback for each type of event and implements its
logic, while Babel handles the events by invoking their appropriate callbacks.
While relatively simple, the event-oriented model provided by Babel allows the
implementation of complex distributed protocols by allowing the developer to
focus almost exclusively on the actual logic of the protocol, with minimal
effort on setting up all the additional operational aspects.

\subsection{Core}  \label{sec:babel:overview:core} The Babel core is the
central component which coordinates the execution of all protocols within the
scope of a process.

As illustrated in Figure \ref{fig:architecture}, every interaction in Babel is
mediated by the core component, as it is this component's responsibility to
deliver events to each protocol's event queue. Whenever a protocol needs to
communicate with another protocol, it is the core that processes and delivers
events exchanged between them. When a message is directed to a protocol in
another process, the core component delivers it to the network channel used by
the protocol, which then sends the message to the target network address. That
message is then handled by the core on the receiving process that ensures its
delivery to the correct protocol.

Besides mediating interaction between protocols (both inter and intra process),
the core also keeps track of timers setup by protocols, and delivers an event to
a protocol whenever a timer setup by the protocol is triggered.

\subsection{Network} \label{babel:overview:network} Babel employs an
abstraction for networking which we name \emph{channels}. Channels abstract all
the complexity of dealing with networking, and each one provides different
behaviors and guarantees. Protocols interact with channels using simple
primitives (\texttt{openConnection}, \texttt{sendMessage}, \texttt{closeConnection}), and receive events
from channels whenever something relevant happens. These events are
channel-specific and are handled by protocols just like any other type of event
(i.e., by registering a callback for each relevant channel event).

For instance, the framework provides a simple \texttt{TCPChannel} which allows
protocols to establish and accept TCP connections to/from other processes. This
channel generates events whenever an outgoing connection is established, fails
to be established, or is disconnected, and also whenever an incoming connection
is established or disconnected. Other examples of provided channels include a
channel with explicit and automatic acknowledgement messages, a channel that
creates one connection for each protocol running in different processes, and a
\texttt{ServerChannel} that does not establish connections, only accepts them,
and its corresponding counter-part, the \texttt{ClientChannel}. We also provide
a TCP-based channel that implements the $\varphi$-accrual failure
detector\,\cite{accrual}, which notifies protocols that registered a callback
whenever another process is suspected.

The framework also allows developers to design their own channels if they need
to enforce some specific behavior or guarantee at the network level for a
protocol to function correctly. Network channels  are implemented using
Netty~\cite{netty}, which is a popular Java networking framework. However, the
typical developer of Babel does not have to interact with Netty directly.

A protocol can use any number of channels, and a channel can be shared by more
than one protocol. In the example of Figure \ref{fig:architecture}, two channels
were instantiated by Babel. Channels within a Babel process are instantiated on
demand by the Babel core when protocols are instantiated. Upon protocol instantiation,
protocols define the channels they will be using, instructing the Babel core to
prepare and instantiate the necessary network channels.

\section{API}  \label{sec:babel:api}

Babel is provided as a Java library. Protocols in Babel are developed by
extending an abstract class - \texttt{GenericProtocol}. This class contains all
the required methods to generate events and register the callbacks to process
received events. Each protocol is identified by a unique identifier, used to
allow other protocols to interact with it. There is also a special \emph{init}
event that protocols must implement, which is usually employed to define a
starting point for the operation of the protocol (e.g., communicate with some
contact node already in the system or setup a timer event).

The API can be divided in three categories: timers, inter-protocol communication
(within the same process), and networking.

\subsection{Timers}
Timers are essential to capture common behaviors of distributed protocols. They
allow the execution of periodic actions (e.g., periodically exchange information
with a peer), or to conduct some action a single time in the future (e.g.,
define a timeout).

In Babel, using timers can be achieved as follows. First, the developer needs to
create a Java class that represents a timer, with a unique type identifier and
extending the generic \texttt{ProtoTimer} class. Additionally, a timer might
have any number of fields or logic as the developer needs.
Listing~\ref{lst:timers} shows an example of the usage of timers in a protocol,
with a timer that contains a counter that can be decremented (lines~$1$-$11$). To
use the timer in a protocol, a callback method must be defined to be executed
once the timer expires. This method must receive as parameters the timer object
and its instance id, which is generated upon setting up the timer
(lines~$21$-$23$). This callback is registered by calling the method
\texttt{registerTimerHandler}, which takes as arguments the unique type
identifier of the timer, and the callback itself (line~$18$). After registering
the handler, any number of single-time or periodic timers can be setup using the
methods \texttt{setupTimer} or \texttt{setupPeriodicTimer}, respectively. These
methods take as parameters an instance of the timer object, and the delay to
trigger the timer (in milliseconds). The periodic timer also receives a third
parameter: the periodicity after the first timeout (line $19$). Cancelling timers
is also possible -- for this, we simply call the method \texttt{cancelTimer}
with the identifier of a previously setup timer as parameter. This identifier is
obtained from the return value of the methods used to setup the timer or from
the second parameter of the callback (line~$22$).

\begin{lstlisting}[style = babel, escapeinside={(*@}{@*)}, %
    caption={Timer Example}, label={lst:timers}]
public class CounterTimer extends ProtoTimer {
    public static final short TIMER_ID = 101;
    int counter;

    public CounterTimer(int initialValue) {
      super(TIMER_ID);
      counter = initialValue;
    }

    public int decCounter() {
      return --counter;
    }
}

public class CountdownProtocol extends GenericProtocol {
    public CountdownProtocol() {
      super("CountdownProtocol", 100);
    }

    @Override
    public void init(Properties props) throws HandlerRegistrationException {
        registerTimerHandler(CounterTimer.TIMER_ID, this::handleCounterTimer);
        setupPeriodicTimer(new CounterTimer(10), 1000, 300);
    }

    private void handleCounterTimer(CounterTimer timer, long timerId) {
        if(timer.decCounter() == 0) cancelTimer(timerId);
    }
}\end{lstlisting}

\subsection{Inter-protocol communication} \label{sec:babel:ipc}

Our framework supports multiple protocols executing concurrently in the same
process. As such, we offer mechanisms for these protocols to interact with each
other, allowing them to cooperate or delegate responsibilities. For instance, we
could create a solution where a message dissemination protocol takes advantage
of a peer-sampling protocol to obtain samples of the system's
filiation\,\cite{hyparview,plumtree,peersamplingframework}.
Listing~\ref{lst:ipc} depicts such an example.

To support this, Babel provides two types of communication primitives:
one-to-one requests/replies and one-to-many notifications. Similarly to timers,
requests, replies, and notifications need to extend a generic class
(\texttt{ProtoReply}, \texttt{ProtoRequest}, and \texttt{ProtoNotification}
respectively) and have unique type identifiers. They can also have any extra
arbitrary state and/or logic. Again, similarly to timers, we need to register a callback for
each type of communication primitive (as seen in lines $7$, $8$, and $25$). The
callbacks all have the same parameters: the object that was sent and the
identifier of the protocol who sent it. In order to send requests and replies,
the methods \texttt{sendRequest} and \texttt{sendReply} are used. These methods
take as parameters the Request or Reply to be sent and the destination protocol
(as seen in lines $9$ and $28$ where, in the latter, the protocol replies to the
sender of the request). For notifications however, the method
\texttt{triggerNotification} (line $31$) does not require a destination,
instead, every protocol that subscribed to that type of notification receives it
(as exemplified in line $7$).

\begin{lstlisting}[style = babel, escapeinside={(*@}{@*)}, %
    caption={Inter-process Communication Example}, label={lst:ipc}]
public class DisseminationProtocol extends GenericProtocol {
    public DisseminationProtocol() {
      super("DisseminationProto", 200);
    }

    @Override
    public void init(Properties props) throws HandlerRegistrationException {
        subscribeNotification(MembershipChangeNot.ID, this::onMembershipChange);
        registerReplyHandler(MembershipReply.ID, this::onMembershipReply);
        sendRequest(new MembershipRequest(), MembershipProtocol.PROTOCOL_ID);
    }

    private void onMembershipChange(MembershipChangeNot not, short emitter) {
        updateMembership(not.getChanges());
    }

    private void onMembershipReply(MembershipReply reply, short from) {
        setMembership(reply.getNodes());
    }
}

public class MembershipProto extends GenericProtocol {
    public static final short PROTOCOL_ID = 300
    public MembershipProto() {
      super("MembershipProto", PROTOCOL_ID);
    }

    @Override
    public void init(Properties props) throws HandlerRegistrationException {
        registerRequestHandler(MembershipRequest.ID, this::onMembershipRequest);
    }

    private void onMembershipRequest(MembershipRequest request, short from) {
        sendReply(new MembershipReply(currentMembership), sourceProto);
    }

    private void onConnectionEstablished(Peer peer){
        triggerNotification(new MembershipChangeNot(peer));
    }
}
\end{lstlisting}

\subsection{Networking} \label{sec:babel:network}

Naturally, as a framework for distributed protocols, Babel also provides
abstractions to deal with networking (including management of connections). For
this, we provide different network channels with different capabilities. The
interaction of protocols with channels is (mostly) similar across different
channels.
Listing~\ref{lst:networking} presents an example of a protocol that sends a ping
message to a random peer, waits to receive a pong response message, and then
chooses a new random peer to repeat the same behavior endlessly.

To use a channel, we start by setting up the properties for that channel. In the
case of the \texttt{TCPChannel}, shown in the example, the required properties
are the binding address and port for the listen socket (lines~$9$~to~$11$);
other channels can consider different properties (e.g., a server channel that
takes as property the maximum number of simultaneous client connections). Next,
we create the channel by calling the method \texttt{createChannel}, passing the
name of the channel and the properties object (line $12$), and receiving an
identifier representing the created channel. This identifier is useful if a
protocol uses multiple channels simultaneously, to be able to select which
channel to use to send a specific message, and to register different callbacks
for different channels. Similarly to timers and requests/replies, we also need
to create a class for each network message to be sent through a channel. Besides
extending a generic class, and having a unique type identifier, the developer
must also define a serializer for each message to enable the message to be
encoded and decoded into network buffers. In lines $49$-$59$ we show an example
of a serializer for a message that only contains a single field. The serializer
could be any function that converts Java objects to/from byte arrays, so one
could use a JSON library or Java's own serializer (that however, would incur in
additional overhead). We register the serializer for each message in a channel
(since we could use different serializers for the same message across different
channels) by using the method \texttt{registerMessageSerializer} (lines
$13$-$14$). We also need to register the callback for when a message is
received, by using the method \texttt{registerMessageHandler} (lines $15$-$16$).
Note that different callbacks for the same message across different channels are
supported. The callbacks to handle received messages take as arguments the
message itself, the sender's IP address and port, the sender's protocol
identifier, and the channel from which the message was received (as shown in
lines $22$ and $25$).

\begin{lstlisting}[style = babel, escapeinside={(*@}{@*)}, %
    caption={Networking Example}, label={lst:networking}]
public class PingPongProto extends GenericProtocol {
    List<Host> peers;

    public PingPongProto()  {
      super("PingPong", 400);
    }

    @Override
    public void init(Properties props) throws HandlerRegistrationException {
        peers = readPeersFromProperties(props);
        Properties channProps = new Properties();
        channProps.setProperty(TCPChannel.ADDRESS_KEY, props.getProperty("addr"));
        channProps.setProperty(TCPChannel.PORT_KEY, props.getProperty("port"));
        int cId = createChannel(TCPChannel.NAME, channProps);
        registerMessageSerializer(cId, PingMsg.ID, PingMsg.serializer);
        registerMessageSerializer(cId, PongMsg.ID, PongMsg.serializer);
        registerMessageHandler(cId, PingMsg.ID, this::uponPing);
        registerMessageHandler(cId, PongMsg.ID, this::uponPong);
        registerChannelEventHandler(cId, OutConnectionDown.EVENT_ID, this::uponOutConnnDown);
        registerChannelEventHandler(cId, OutConnectionFailed.EVENT_ID, this::uponOutConnFailed);
        registerChannelEventHandler(cId, OutConnectionUp.EVENT_ID, this::uponOutConnUp);
        openConnection(chooseRandomPeer(peers));
    }

    private void uponPing(PingMsg msg,Host from,short sourceProto,int cId) {
        sendMessage(new PongMsg(msg.getValue()), from, TCPChannel.CONNECTION_IN);
    }

    private void uponPong(PongMsg msg,Host from,short sourceProto,int cId) {
        closeConnection(from);
        openConnection(chooseRandomPeer(peers));
    }

    private void uponOutConnUp(OutConnectionUp ev, int cId) {
        sendMessage(new PingMsg(System.currentTimeInMillis()), ev.getPeer());
    }

    private void uponOutConnFailed(OutConnectionFailed<ProtoMessage> ev, int cId) {
        openConnection(chooseRandomPeer(peers));
    }

    private void uponOutConnDown(OutConnectionDown ev, int channelId) {
        openConnection(chooseRandomPeer(peers));
    }
}

public class PingMsg extends ProtoMessage {
    public final static short MSG_ID = 401;
    private final long timestamp;

    public PingMsg(long timestamp) {
        super(MSG_ID);
        this.timestamp = timestamp;
    }

    public int getTimestamp() {
        return timestamp;
    }

    public static ISerializer<PingMsg> serializer = new ISerializer<>() {
        @Override
        public void serialize(PingMsg msg, ByteBuf out) throws IOException {
            out.writeLong(msg.timestamp);
        }
        @Override
        public PingMsg deserialize(ByteBuf in) throws IOException {
            long timestamp = in.readLong();
            return new PingMsg(timestamp);
        }
    };
}
\end{lstlisting}

A message can be sent using the \texttt{sendMessage} method, which takes as
arguments the message to be sent, the destination address/port, the channel
identifier (if more than one channel is being used) and optionally, the
destination protocol. An additional parameter representing the connection to use
can be passed. The interpretation of this parameter is however, channel
dependent. Finally, each channel is responsible for generating notifications for
relevant events that occur in it, for which the protocol can register callbacks.
In the given example, the events generated by the \texttt{TCPChannel} (described
before) are registered using the method \texttt{registerChannelEventHandler} (as
seen in lines $17$-$19$).

\section{Evaluation}\label{sec:casestudies}

Our evaluation is based on two case studies of popular distributed applications.
Our first case study is a simple peer-to-peer (P2P) application that
disseminates messages to the network by leveraging two protocols, the
HyParView\,\cite{hyparview,leitao:msc} membership protocol that builds an unstructured
overlay and provides nodes with stable neighboring nodes; and a Flood
Dissemination protocol that disseminates a message to all neighboring nodes. Our
second case study is a simple state machine replication (SMR) application, that
leverages a MultiPaxos\,\cite{multipaxos} consensus protocol to order the operations
of an in-memory key-value store.

Our evaluation is divided in three parts. In Section~\ref{subsec:imp} we provide
implementation details over the protocols (implemented with Babel) for each
case study. In Section~\ref{subsec:code}, we perform a code review of publicly
available MultiPaxos protocols and provide a comparison with our MultiPaxos
implementation with Babel. Finally, in Section~\ref{subsec:eval} we present a
performance evaluation on both case study applications.

\subsection{Implementation Details}\label{subsec:imp}

In this section we provide some implementation details over the
protocols implemented with Babel for each case study. Each case study was implemented by
a single PhD student familiarized with the protocols, and took about a work day to
implement, analyze, and perform simple experiments to verify their operation and
functionality.

\paragraph{P2P Application}

The P2P application leverages two protocols (HyParView and a Flood
Dissemination) that interact with each other through\,\cite{leitao:msc} the inter-protocol
communication mechanisms of Babel. In more detail, the Flood protocol registers
interest in \emph{neighbor up} and \emph{neighbor down} notifications that are
produced by the HyParView protocol whenever it connects to a new neighboring
node, or disconnects from a neighboring node, respectively.

The HyParView protocol is implemented with $359$ lines of Java code. This
includes $15$ event handlers, of which, $8$ are network message handlers, one
for each message processed by the protocol; $2$ are timer handlers, one for a
periodic action, and another for a timeout action; and $5$ are channel event
handlers, to the \texttt{TCPChannel} events. The events that support the operation of the
protocol are implement with $361$ lines of code, of which the majority ($306$)
is dedicated to message events. Furthermore, a data structure implemented with
$108$ lines of code maintains the state of the protocol. In total, our
implementation has $828$ lines of code.

The Flood dissemination protocol is implemented with $74$ lines of code. This
includes $5$ event handlers, of which $1$ is a message handler that processes
the broadcasted messages; $1$ is a request handler that processes the request of
the application to broadcast a message; and $2$ are notification handlers that
process the notifications triggered by HyParView. The protocol has $3$ events
that support its operation (a \texttt{BroadcastMessage}, a
\texttt{BroadcastRequest}, and a \texttt{DeliveryReply}) that are implemented with $91$
lines of code. We have also implemented a utility class that
produces numeric hashes of the contents of the broadcasted messages to serve as
unique message identifiers.

\paragraph{SMR Application}

The SMR application is an in-memory key-value store that submits (client issued) operations to a
consensus protocol for ordering. The consensus protocol exposes a notification
that notifies the application of the execution order of operations. We
have implemented two variants of MultiPaxos: a classic MultiPaxos variant where
the acceptors inform all learners of their accepted values, and; a distinguished
learner MultiPaxos variant, where the acceptors inform the leader (i.e., current
proposer) of the accepted value, who then informs all learners.

The classic MultiPaxos is implemented with $235$ lines of code, this includes
the implementation of the proposer, acceptor, and learner components of the
protocol. The protocol has $13$ event handlers, where $4$ are network message
handlers, one for each protocol message; $3$ are timer handlers for handling
different timeout operations of the protocol (e.g., suspect the leader is dead,
try to reconnect to a peer, enforce leadership); $1$ is a request handler that
handles the interface with the application(s); and finally, $5$ are channel
event handlers for handling the \texttt{TCPChannel} events. The events of the
protocol are implemented with $250$ lines of code, where $163$ lines of code are
for messages. Furthermore, we have implemented some utility classes that help
maintain the state of each consensus instance and operation, which were
implemented with $250$ lines of code. In total, our implementation has $735$ lines of code.

The distinguished learner MultiPaxos is implemented with $247$ lines of code.
This implementation is very similar to the previous, with one additional network
message handler for the message used to inform the learners of
the accepted value.

\subsection{Code Review}\label{subsec:code}

\begin{table}[t]
\centering
\footnotesize
\caption{MultiPaxos Implementations Lines of Code}
\begin{tabular}{ l 
                              | l 
                                          |  r 
  }
\toprule
Implementation & Component & Java Lines of Code \\
\midrule

\multirow{4}{*}{Babel-MultiPaxos-Classic} & \bfseries Total & \bfseries $735$ \\
                                          & Main Protocol Logic & $235$ \\
                                          & Events & $250$ \\
                                          & Utils & $250$ \\

\midrule
\multirow{4}{*}{Babel-MultiPaxos-DistLearner} & \bfseries Total & \bfseries $787$ \\
                                          & Main Protocol Logic & $247$ \\
                                          & Events & $290$ \\
                                          & Utils & $250$ \\

\midrule
\multirow{5}{*}{MyPaxos}          & \bfseries Total & \bfseries $1814$ \\
                                  & Main Protocol Logic & $683$ \\
                                  & Messaging & $306$ \\
                                  & Utils & $487$ \\
                                  & Other & $338$ \\

\midrule
\multirow{7}{*}{WPaxos}           & \bfseries Total & \bfseries $22909$ \\
                                  & Main Protocol Logic & $2283$ \\
                                  & Messaging & $862$ ($+$ $78$ Protobuf) \\
                                  & Networking & $1376$ \\
                                  & Support & $536$ \\
                                  & Utils & $497$ \\
                                  & Other & $17891$ \\

\bottomrule
\end{tabular}
\label{tab:code}
\end{table}

In this section we provide a code review of publicly available MultiPaxos Java
implementations, along with a comparison with our Babel MultiPaxos
implementations. We chose the two MultiPaxos Java implementations on GitHub
with the highest number of stars to perform this evaluation. The
implementations are \emph{MyPaxos}\,\cite{mypaxos} and
\emph{WPaxos}\,\cite{wpaxos}, which had $131$ and $130$ stars at the time,
respectively. In fact, these two implementations provide an interesting
comparison as they, while implementing very similar MultiPaxos variants, have
key differences that highlight some of the benefits of Babel. MyPaxos presents a
very simple but naive implementation, while WPaxos presents a more complex and
performant implementation. Despite this, they share the following design
choices: $i)$ the roles of proposer, acceptor, and learner are logically split
across different Java classes; $ii)$ instances are not executed concurrently (i.e., an
instance only starts after the previous one is decided); and $iii)$ every node
that receives client operations competes to be the leader, meaning the
application needs to redirect all operations to a single node to avoid poor
performance and/or failed operations.

The only considerable design difference between the two implementations is that MyPaxos
uses the classic message flow of MultiPaxos, while WPaxos uses the distinguished
learner variant. In the following, we detail implementation aspects that were found
while reviewing the code of each solution. Furthermore, Table~\ref{tab:code}
summarizes the number of lines of code for each implementation, where it is
noticeable that our implementations have significant less lines of code.

\paragraph{MyPaxos}
This implementation sacrifices performance and optimizations in favor of a
small, simple-to-understand, and clear code structure. In short, in this
implementation each node has three main threads, each corresponding to one of the
Paxos roles. These communicate by sending messages (both to other nodes and to
other roles within the same process). The implementation uses a simple
networking library and relies on Java's native object serialization to encode
messages before sending them.

One thing we noted in this implementation, is that learners would periodically
request learned values from other learners, to know when an instance had been
decided. However, this heavily limited the rate at which operations were
executed. As such, we made a small modification to the protocol, making learners
eagerly broadcast their values as soon as they were accepted. This
modification was done with about $10$ lines of code without changing the
implemented logic by the original authors.

Additionally, we observed that this implementation was not very robust when
tested under heavy load. The first issue we faced was a concurrency issue
resulting in multiple threads accessing the same data structure, in this case a
\texttt{HashMap}, which we solved by changing the data structure to the
thread-safe \texttt{ConcurrentHashMap}. The second issue, resulted from a
simplification made in the implementation that would spawn a new thread for each
Paxos instance that would monitor a timeout. However, when executing the
protocol under heavy load for a long period of time, the JVM crashed due to
being unable to allocate memory for new threads. Babel aims at shielding the
developer from issues such as these. In fact, Babel avoids concurrency
issues by having sequential execution of events within a protocol, and its timer
management allows protocols to register timers without the overhead of creating
and managing extra threads.

In comparison with our Babel implementations, MyPaxos while being relatively
simple, still requires more lines of code to implement than any of our
Babel MultiPaxos implementations. This is mainly due to MyPaxos having to
deal with the execution support code (e.g., threading, messaging, timers)
that is handled by Babel.

\paragraph{WPaxos}
This implementation is much more robust and performant than MyPaxos. It uses a
custom communication layer, leveraging on the Netty\,\cite{netty} network
framework, using TCP connections for large messages and UDP packets for small
messages. Moreover, it serializes messages using Protocol Buffers\,\cite{protobufs}. While
it also splits the Paxos roles in different Java classes, it handles the
interactions between roles within the same process more efficiently than
MyPaxos, by using direct method invocation instead of sending network messages.
We were able to deploy and test this implementation without requiring any
modification.

In comparison with our Babel implementations, it is clear that the use of Babel
can greatly reduce the effort of programmers in creating working prototypes of
distributed applications. While a considerable number of lines deal with
features not implemented in our prototypes (mostly the $16275$ lines of
\emph{Other} reported in Table~\ref{tab:code}) such as persistency, there is
still a substantial difference in the number of lines for the main protocol
logic. This difference comes mostly from the logic dedicated to creating and
handling timers, multiplexing received messages and triggered timers, as well as
making sure received messages are handled sequentially. We note that using Babel
all of these aspects do not have to be handled by the developer.

\subsection{Performance Evaluation}\label{subsec:eval}

\begin{figure}[t]
    \centering
    \subcaptionbox{Reliability of P2P application.\label{fig:reliability}}{
      \includegraphics[width=0.48\linewidth]{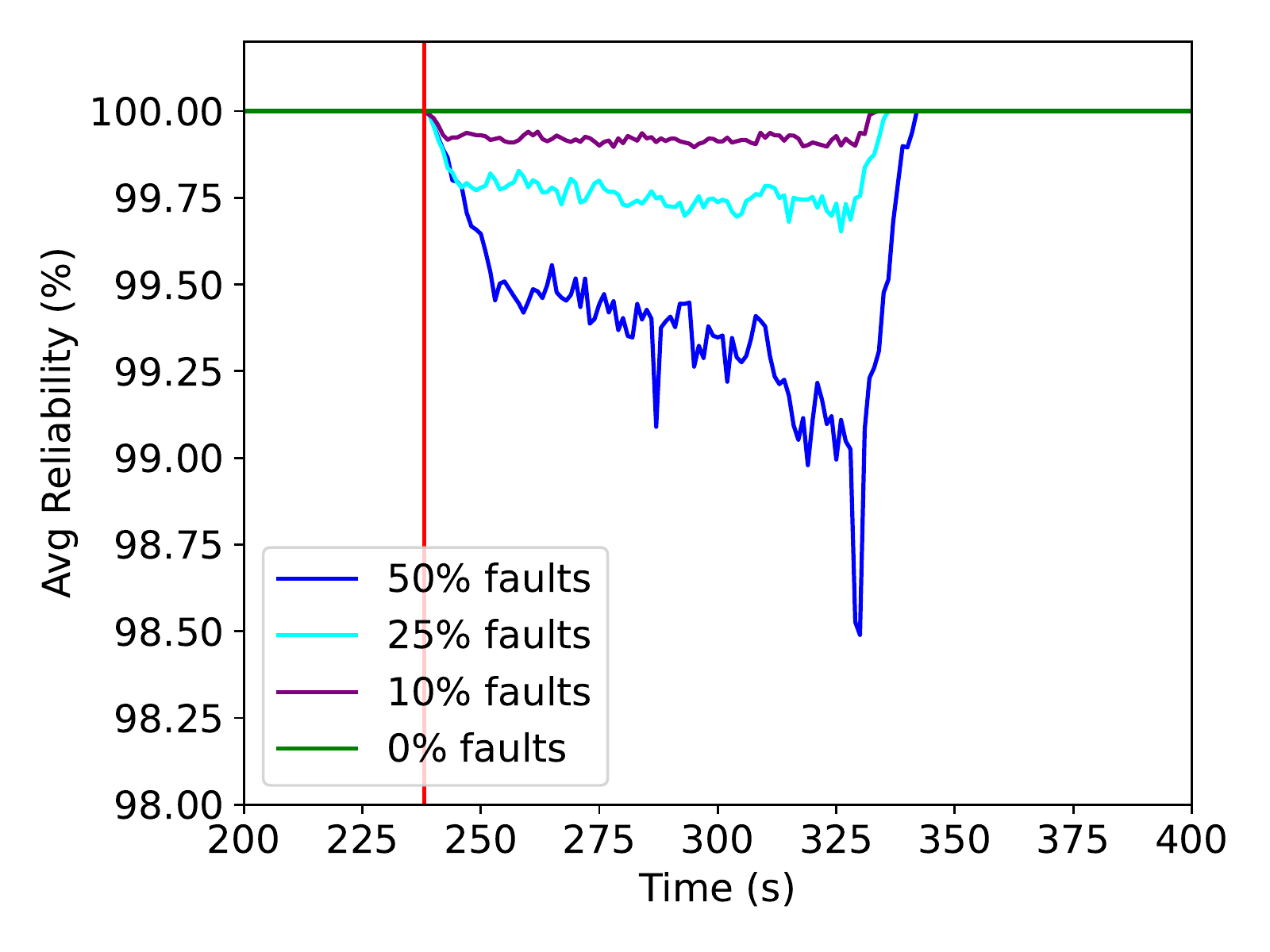}}
    \subcaptionbox{Throughput of Multipaxos implementations.\label{fig:throughput}}{
      \includegraphics[width=0.48\linewidth]{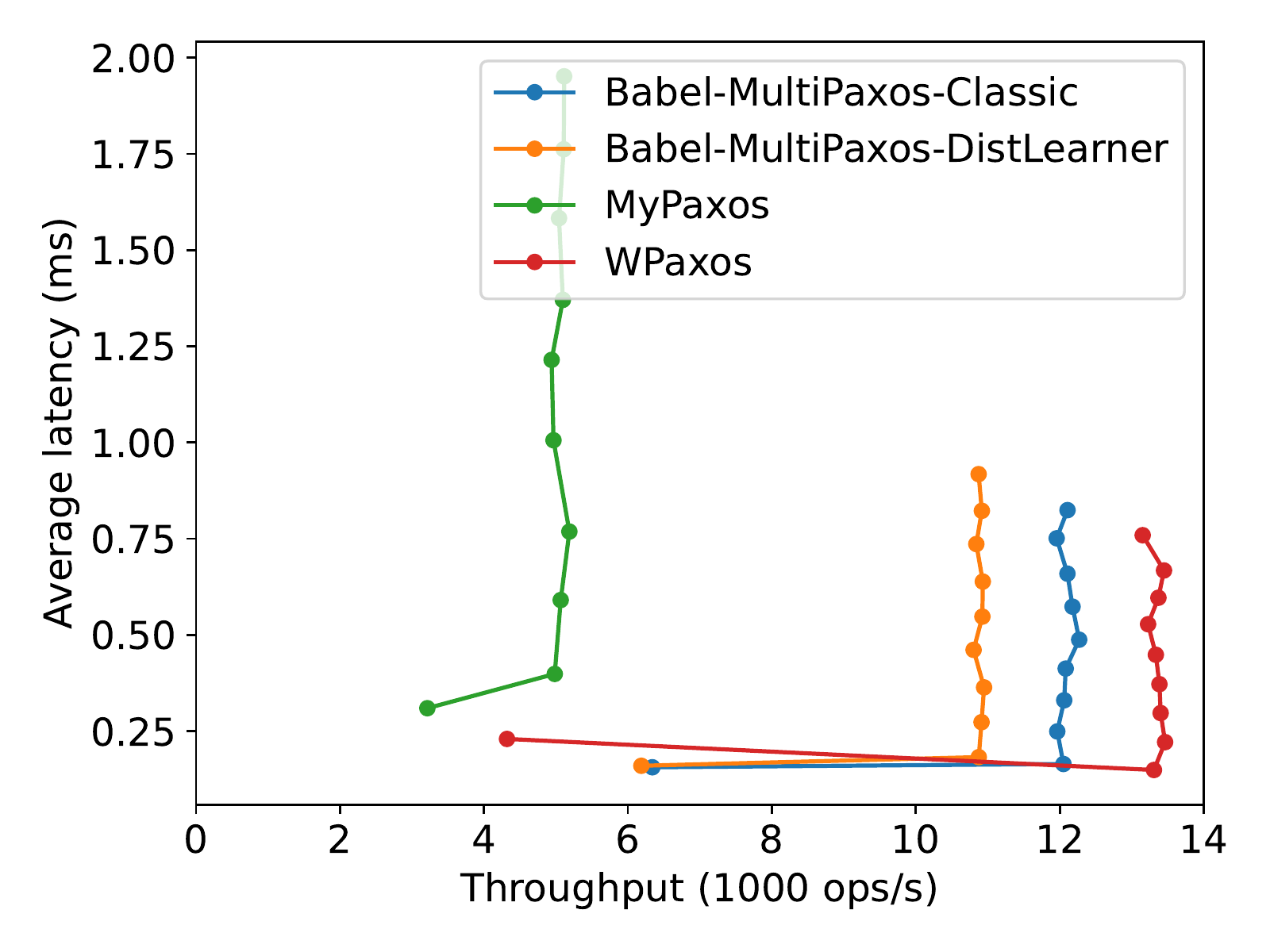}}
  \caption{Performance evaluation results
  \label{fig:performance}}
\end{figure}

In this section we describe our performance evaluation using the two case study
applications. Our goal was, on the one hand to verify the correctness of our
implementations and how they behave under faults (P2P application), and on the
other hand to provide insights about the relative performance of these
implementations when compared with standalone ones (SMR application). In
the following we describe the experimental setup and discuss the performance
results for each case study.

\paragraph{P2P Application} Our experiments with the P2P application showcase a
typical P2P evaluation with a network with a considerable number of nodes. In our
experiments we measure the average reliability of message delivery in four
different scenarios: $i)$ a fault-free scenario; and fault scenarios where different
fractions of nodes fail simultaneously $ii)$ $10\%$ faults; $iii)$ $25\%$ faults; and $iv)$ $50\%$ faults.

To execute our experiments we execute $100$ docker containers  distributed
evenly across $4$ identical servers with an AMD EPYC 7281 CPU and 128GiB of
memory. Each container executes an instance of the application. We have applied
latency among the containers with the Linux \texttt{tc} tool, based on a network
generated with inet\,\cite{inet}, having a mean latency of $293.39$
milliseconds, to have a more realistic scenario. Each experiment runs for $12$
minutes. Nodes start disseminating messages with a ratio of one message per
second after two minutes since the start of the experiment until they have
broadcasted a total of $500$ messages. Faults are introduced at the middle of
the experiment. Each experiment was executed three times. Results show the
average of all runs for each experiment.

Figure~\ref{fig:reliability} presents the results of all scenarios, with a
vertical red line indicating the point of the experiments where faults were
introduced. The x axis represents the time of the experiment when messages were
being disseminated. The y axis represents the average reliability of message
delivery. As one can see, our experiments show the expected results for a flood
strategy dissemination protocol on top of a self-healing stable overlay that is
able to quickly recover, having a very low reliability drop in scenarios with
faults (not reaching bellow $98\%$ reliability).

\paragraph{SMR Application}

Our experiments with the SMR application present an experimental
comparison of different MultiPaxos implementations. In more detail, we compare
our MultiPaxos implementations with the public ones described previously.
To make comparisons fair, we adapted our key-value store application to work
with MyPaxos and WPaxos. Furthermore, we disabled batching from all solutions,
disabled the persistency in MyPaxos and configured WPaxos to perform the
persistency asynchronously.

These experiments are executed in $4$ identical servers with an Intel Xeon Gold
5220 CPU and 96GiB of memory, hosted in the Grid5000\,\cite{grid} platform.
Three servers execute the replicas of our application, while the last one
executes the client. The client executes the YSCB\,\cite{ycsb} benchmarking
tool, issuing operations in a closed loop to the key-value store application
(which then orders them using the one of the studied MultiPaxos implementations)
with a $50\%$ read/write ratio. We vary the number of client threads from $1$
to $10$, with increments of $1$. Each
experiment executes $5$ times for a period of $90$ seconds. We only consider
data points from the $30$ second mark to the $80$ second mark to eliminate
possible instabilities at the start or the end of an experiment. Results show
the averages of each run for each implementation.

Figure~\ref{fig:throughput} presents the average throughput (in the x axis) against
the average latency (in the y axis) of operations for each implementation.
We note that MyPaxos is the implementation with the least throughput. This is
most likely due to the serialization used by the implementation which is known
to have non-negligible overhead. In contrast, Babel MultiPaxos implementations
show comparable performance results with the significantly more complex WPaxos implementation,
which is able to achieve higher throughput due to the optimizations performed in
the code albeit, at a much higher development overhead. We note that we validated
the correctness of the solutions by inspecting the state of the replicated in-memory key-value store
for each solution.

\section{Conclusion}\label{sec:end}

In this paper we propose Babel, a Java framework to support the quick
prototyping of distributed algorithms, particularly those that focus on
fault-tolerance and dependability aspects of distributed systems. This tool can
be used by researchers, practitioners, and teachers of distributed systems
courses. Babel provides a set of abstractions that allows the developer to focus
their efforts on implementing the algorithm logic, following an API that is
close to the typical presentations either by descriptions or pseudo-code of such
algorithms. Furthermore, Babel shields the developer from dealing with low-level
networking aspects, as well as concurrency issues that arise when implementing
protocols in a naive way, as shown in our comparison with public implementations
of Paxos. We illustrated the benefits of Babel through two different case
studies, and conducted a small experimental validation that shows that
implementations that benefit from Babel present competitive performance when
considering significantly more complex stand-alone implementations.

As future work, we plan to continue enriching Babel with novel network channels
that can provide debugging tools that inspect the state of executing protocols
and systems. We also plan on continuing to work on simpler and more powerful
abstractions and interfaces to simplify the development of Babel protocols,
along with integrated logging tools to facilitate debugging.

\bibliographystyle{IEEEtran}
\bibliography{bib}

\end{document}